\documentclass[letterpaper,twoside,twocolumn,english,prl]{revtex4}
\usepackage[T1]{fontenc}
\usepackage[latin9]{inputenc}
\usepackage{amsmath,xcolor}
\usepackage{esint}
\usepackage{graphicx}
\usepackage{placeins}
\usepackage{float}
\usepackage{subfigure}

\makeatletter

\@ifundefined{textcolor}{}
{%
 \definecolor{BLACK}{gray}{0}
 \definecolor{WHITE}{gray}{1}
 \definecolor{RED}{rgb}{1,0,0}
 \definecolor{GREEN}{rgb}{0,1,0}
 \definecolor{black}{rgb}{0,0,1}
 \definecolor{CYAN}{cmyk}{1,0,0,0}
 \definecolor{MAGENTA}{cmyk}{0,1,0,0}
 \definecolor{YELLOW}{cmyk}{0,0,1,0}
}

\usepackage{babel}

\makeatother

\usepackage{babel}
\begin{document}

\title{Human wealth evolution: trends and fluctuations}

\author{Paolo Sibani$^{1,\ast}$ and Steen Rasmussen$^{1,2\ast}$}
\affiliation{$^{1,\ast}$Center for Fundamental Living Technology (FLinT) \\
Department for Physics, Chemistry and Pharmacy, University of Southern Denmark\\ 
 $^{2}$Santa Fe Institute, Santa Fe, NM 87501, USA \\
$^\ast$ paolo.sibani@sdu.dk  and steen@sdu.dk}
\begin{abstract}
Is a causal description of human wealth history conceivable? 
To investigate the matter we introduce a simple causal albeit strongly aggregated model, assuming that the observed wealth growth is mainly
 driven by human collaborative efforts whose intensity itself increases with increasing wealth. 
As an empirical reference we use time series describing 
eight centuries of per capita annual gross domestic products (GDP) of three European countries, the UK, France and Sweden.
The model requires a population large enough  for  disruptive events, e.g. famine, epidemics and wars, 
 not to destroy the fundamental workings of society. Cultural interchanges between different geographical areas are not explicitly taken into account.  
The wealth evolution trend can then be described by an ordinary differential equation with three free parameters: 
one producing a short term exponential growth rate, one defining  an additional  minute constant growth rate and finally one that
 specifies the  time scale below which  human collaboration intensity  can be treated as  constant. 
 Beyond that scale, wealth enhances the fundamental collaborative infrastructure of society. 
The solution features a finite time singularity, which implies infinite GDP per capita in finite time and thus suggests a lack of long term sustainability.
The year at which the singularity occurs has a slight variation near  2020 AD
 from one country to another. 
GDP time series curtailed after 1900 AD produce similar values for the occurrence of the singularity, which thus could be predicted more than hundred years ago.
Curtailing the GDP series from the early years up to 1700 AD also produces stable and consistent predictions for the singularity time.
{Power spectra are obtained for de-trended data spanning eight centuries, as well as for the first and  last four centuries
of the same period.
All spectra have an overall signature where the power decays as the inverse frequency  squared.
The embedded peaks  are reminiscent of the  cycles described in the economic literature, 
but are  also present in time series far predating industrialization.  
The  background fluctuations in the GDP series is tentatively  interpreted as  societal  response
 to disruptive stochastic events. e.g. new economic activities following epochal discoveries, as well as  wars and epidemics.}
\end{abstract}
\maketitle

\section{Introduction}
The use and availability through history of resources, natural as well as financial, was recently reviewed by Turner~\cite{Turner14}, who highlights a lack of sustainability in the way human activities are presently organized, reinforcing many previous findings, see e.g. Meadows et al. \cite{MedowsEtAl1972}. 
A number of additional indicators point to an ongoing societal transformation linked to the way new technologies co-evolve with social structures \cite{GrowingGap_2018}. 

In this work we reach similar conclusions, but base them on a different line of arguments, rooted in the way human interactions
 generate the economic output measured by Gross National Product (GDP) per capita~\cite{Costanza14}.
GDP measures the market value of products and services and is a dubious  measure of the quality of life~\cite{Stiglitz10}. 
More importantly, it is not directly linked to the abundance or scarcity of any particular resource.
With population growth factored out, an increasing GDP per capita reflects the evolution of a societal  organization,
specifically of  the physical infrastructures and  institutions needed to maintain a population with  shared 
 cultural values and social and  technological know-how. The structural framework needed to generate
 the   economic activities measured by
per capita GDP will in the following  be called `wealth'. The latter
 is  strongly linked to the task selection and labour 
 organization that generate market value using whichever resources are available.
It  includes far more than  tradable assets,  and  its evolution generates the 
complex dynamics observed in 
 economics~\cite{Beinhocker06}.

Our model's predicted trend is in agreement with historical data, but features a finite time singularity, meaning that the GDP per capita will soon diverge to infinity. 
That historical trends extracted from economical and demographical indices might feature a singular point 
has been mooted in the literature. 
 Kapitza~\cite{Kapitza1996} showed that the history of   human population size 
  is well  described assuming a growth rate  proportional to the square of the population.
  His hyperbolic growth model is similar to. ours and 
   implies a  finite time singularity, happening very close to our present times.
Johansen and Sornette~\cite{Johansen01}
 analyzed world population and GDP data over two millennia, together with financial time series over more than one century, and  fitted  their data in different, but related ways, the simplest being a power law that diverges at a critical time $t_c$.
 Finite time singularities are thus naturally embedded in key data sets and that rapid changes in the mechanisms producing them
 will occur  to avert  singular behavior.

\begin{figure}[t!]
	\vspace{-3.2cm}
	 \includegraphics[width=0.95\linewidth]{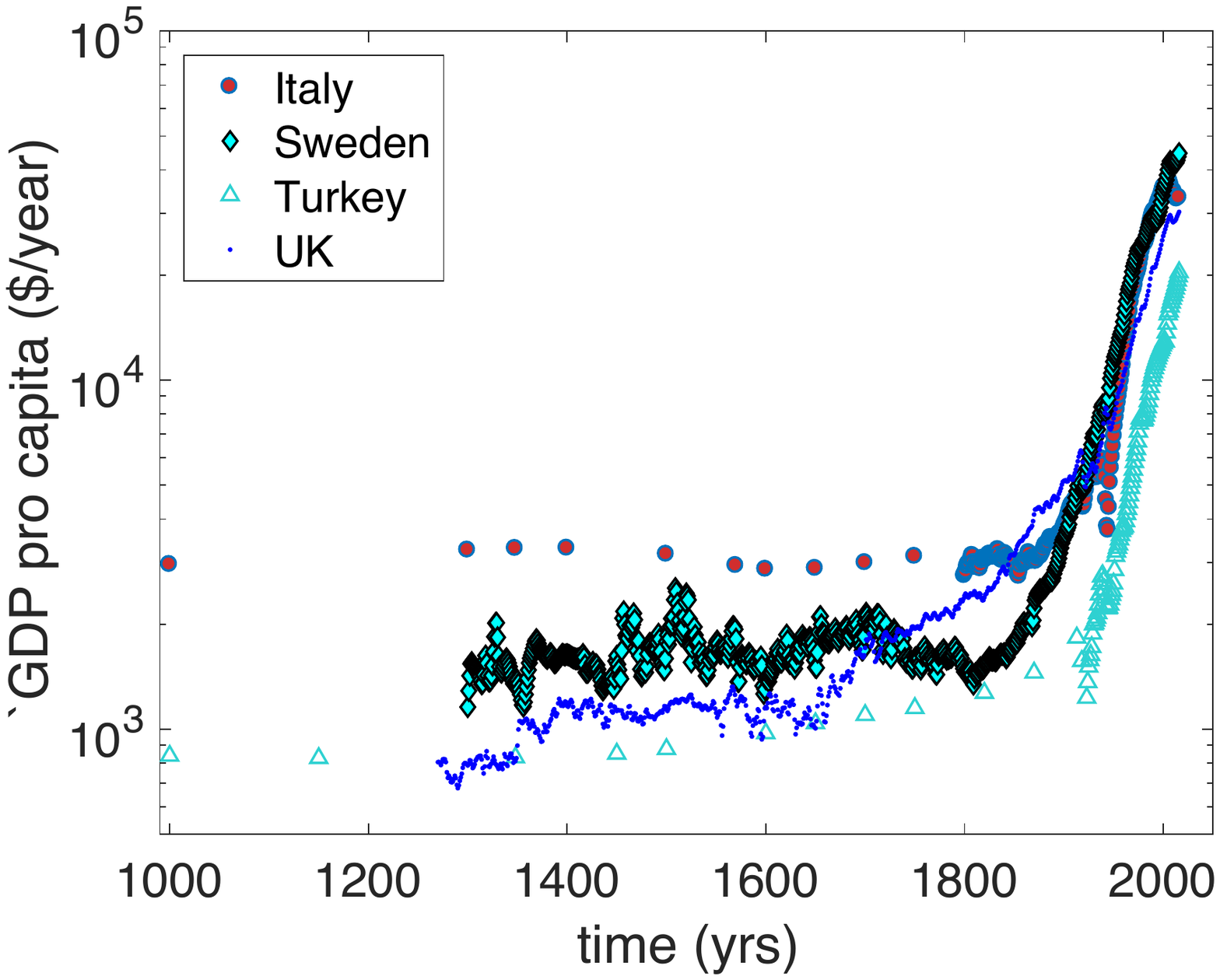}
	\vspace{-3cm}
	\caption{GDP per capita of four different geographical areas plotted versus time on a logarithmic vertical scale.
	Note the large initial differences and the faster than exponential increase occurring in the last two centuries. 
	Data for Italy, Sweden and Turkey from the Maddison data base~\cite{Maddison18}. The UK data are from~\cite{OWD}.
	}
\label{fig:wealth4}
\end{figure}

The kinship of diverse evolving systems, living or not, was recently~\cite{RasmussenSibani2019} discussed with focus on whether their rate of change is decreasing or increasing. 
One type of evolutionary dynamic optimizes a target on fast timescales, e.g. all spontaneous physical processes mini\-mize their free energy and bacterial monocultures optimize fitness in a constant environment~\cite{Lenski94}. 
Evolutionary progress can here be viewed as overcoming a series of ever increasing `record sized' dynamical barriers of system specific sort~\cite{Anderson04,Robe16}.
Evolutionary expansion, the less common type of evolution is not as well understood~\cite{Packard_etal_2019}. 
An example of evolutionary expansion dynamics is given in~\cite{RasmussenSibani2019} based on a time series of British GDP values per person per year~\cite{OWD}, a quantity whose trend has kept increasing faster than exponentially since the late Middle Ages.
Similarly fast GDP growth in different countries over nearly a millennium is shown for convenience in Fig.~\ref{fig:wealth4}.
The same quantity is analyzed in \cite{RosenlystSiboniRasmussen2018} for 28 countries from the onset of the Industrial Revolution onwards and in \cite{RasmussenMosekildeHolst1989} for Danish GDP data only.
In most cases, the growth law from the onset of the Industrial Revolution is well fitted by $f(t) \propto \exp(a  t^2)$, where $t$ is time and $a$ is a proportionality constant.

In this work, GDP time series for France, Sweden and the UK are used as proxy for wealth produced by cultural evolution.
Surprisingly to us, a simple ordinary differential equation (ODE) approximates GDP trends which cover developments from the late middle ages to present times. 
The model assumes that wealth grows at a rate proportional to both \emph{(i)} existing wealth representing both current activity and heritage from the past, and \emph{(ii)} the intensity of human interactions.
Our ODE features a finite time singularity, which suggests a simple endogenous mechanism signalling the end of one evolutionary epoch and the start of a new one.
 
Data streams systematically curtailed to end in $\sim1900$ AD, i.e. earlier than a century ago,  produce GDP trends diverging at times which are all similar and quite close to our near future.
A similar procedure curtailing the beginning of the GDP time series confirms that  the predicted date of the  singularity is robust to input variations, as long as the data stream starts around 1700 AD.
This all implies that a finite time singularity was already predictable long ago and suggests that more recent developments also pointing to imminent changes of socio-technical structures may be consequences of human interaction patterns well established through the history of Western societies.
Our causal model of wealth development assumes that the main circumstances underpinning wealth growth have not changed significantly over hundreds of  years.
This strong assumption is further scrutinized in Section 3.

 {The de-trended data are treated as a stationary process and nine power-spectra are generated, three for each country, based on the
 full time  series and on two curtailed ones, spanning 
the first and  last four centuries of data.
A number of peaks are found, usually explained as cyclic economic phenomena based on modern data  only, see however~\cite{Modelski96}.
Interestingly, all our nine data sets produce similar groups of  oscillation periods. 
More significantly, the background signal falls off as the second inverse power of frequency, indicating the presence of a noisy component with a correlation time of the order of sixty years.
This is tentatively interpreted as due to historical incidents, e.g. epochal innovations, epidemics and wars may, which both excite autonomous oscillations
 and have a lingering effect on GDP, the latter producing the background signal of the power spectra.
}

\section{The model}
Over millennia, human societies have evolved from small isolated groups bound by oral traditions to much larger entities, where interactions are easier, more intense and structured by a common language and norm setting institutions which transcend human lives and secure continuity over manifold generations.
Increasingly faster and more intense communication has underpinned the coordination and collaboration efforts driving human achievements.  

For the last millennium  one may assume, as we presently do, that human agents
 have  always been available to fill in the different functional or `professional' niches of their society and that the size of a population
  has not limited the pace of cultural evolution. 
The bubonic plague (1334 AD) is a relevant example. 
It killed almost half of the European population, but did not destroy social structures and institutions and might even have triggered the wave of cultural innovation that soon after swept through Europe during the Renaissance.

Fig.~\ref{fig:wealth4} reveals considerable wealth differences across countries, or, more precisely, across geographical areas conveniently labeled with their modern names.
Notably, the UK GDP was initially lowest, but overcame all others in the early nineteen century and was again surpassed by the Swedish GDP in the twentieth century.

The difference between evolutionary expansion and evolutionary optimization partly lies in which independent `time' variable best  fits their description. 
Evolution processes promoted by events uniformly distributed in astronomical or `wall-clock' time, e.g. random energy fluctuations in an aging spin-glass, cosmic rays, or genetic mutations in biological organisms, have wall-clock time as their natural variable.
In contrast, human culture is mainly controlled by personal interactions whose intensity has increased dramatically over time.

Limiting ourselves to communication technology in the last eight hundred years in the Westerrn world,, writing, an arduous undertaking for monks in the middle ages, 
was greatly boosted by the introduction  of the printing press. 
Later came the telegraph, the telephone, then radio and tv, and more recently  the internet, hand held devices and the Internet of Things. 
Clearly, the amount of human interactions occurring in, say, one hour of wall-clock time, had a  manyfold  increase over the last eight centuries~\cite{WeAreSocial2018}.

To describe GDP evolution trends we use a dimensionless quantity $w$ called `wealth', which we compare with the likewise dimensionless ratios of GDP time series with their initial values.
Wealth has a simple dependence on its natural time variable, $\tau$, which only describes evolution over short scales. i.e. a few years. 
The dependence on wall-clock time $t$ describes evolution over centuries and includes the effect of an increasing interaction intensity.

Again, our wealth variable is meant to cover the social structures ensuring a society's cohesion and the distribution of knowledge to its inhabitants, e.g. political, religious and education systems, including science and technology. 
Wealth also includes physical infrastructures, e.g. roads, railways and ports, and technical skills, i.e. the ability to build and ma\-nu\-facture all complex items in use, from cathedrals to airplanes.
In other words both the prerequisites for and the products of human economic activities are included.

In terms of our `natural' time variable $\tau$, we posit that wealth per capita is controlled by the simple ordinary differential equation
 \begin{equation}
    \frac{dw}{d\tau} =\alpha w(\tau) +\beta.
\label{def0}
\end{equation}
where $\alpha$ and $\beta$ are positive real  parameters.  
In a system with constant interaction intensity and initial wealth  $w_0$, wealth will thus grow as 
 \begin{equation}
    w(\tau) = \frac{\beta}{\alpha }\left( e^{\alpha \tau} -1
    \right)  +w_0e^{\alpha \tau}  \approx \beta \tau +w_0 \; {\rm for }  \;  \alpha \tau \ll 1.
 \label{tausol}
 \end{equation}
For small values of $\alpha$ we see a slow linear increase proportional to $\beta$, which mirrors the absence of a mechanism destroying wealth once it is created.

The connection between `natural'  and wall-clock time variables were discussed in~\cite{RasmussenSibani2019} for different examples.
For human wealth evolution we here  assume that
 \begin{equation}
 	\tau(t)= \gamma \int_0^t w(t') dt' 
\label{naturalT}
\end{equation}	
which is mathematically equivalently to
   \begin{equation}
       	\frac{d \tau}{dt}= \gamma w(t),
\label{dtau/dt}
\end{equation}
where $\gamma$ is a proportionality constant. 
The connection between $\tau$ and $t$ describes that some fraction $\gamma$ of the wealth produced is used to improve communication speed and efficiency,  be it by e.g. building better roads or faster computer chips. 

Equation (\ref{naturalT}) expresses the current natural time $\tau$ in terms of  the current level of human interactions, which is assumed proportional to the human wealth accumulated over time.  
Expressed equivalently in Equation (\ref{dtau/dt}),
 the current increase in human interactions is proportional to the current wealth. 
For constant wealth, $\tau$ and $t$ are simply proportional. 
However, if  wealth grows, so does the intensity of human interactions, and $\tau(t)$ increases faster than linearly.

To transform Eq.~\eqref{def0} from `natural' to `wall clock' time,
we substitute $\frac{dw}{d\tau}$ from Eq.~\eqref{def0} and $\frac{d\tau}{dt}$ from Eq.~\eqref{dtau/dt} in 
\begin{equation}
	\frac{dw(t)}{dt}= \frac{dw}{d\tau} \frac{d\tau}{dt},
\label{dw/dt}
\end{equation}
 to obtain
\begin{equation}
	\frac{dw(t)}{dt}= (\alpha w(t) + \beta) w(t) = \alpha w^2(t) + \beta w(t).
\label{fullEq}
\end{equation}
The proportionality constant $\gamma$  is absorbed in the fitted parameters $\alpha$ and $\beta$, and is  left unspecified.

The solution to Eq.~\eqref{fullEq},
 \begin{equation}
 	w(t)  = \frac{\beta w_0 }{e^{-\beta  t}(\beta+\alpha w_0 )-\alpha w_0 },
 \label{sol4}
 \end{equation}
  features a finite time singularity at 
 \begin{equation}
 	t^*=\frac{1}{\beta}\log \left({1+\frac{\beta}{\alpha w_0}}
   	\right).
 \label{fts}
 \end{equation} 
The singularity occurs at $t^*=\frac{1}{\alpha w_0}$ in the limit $\beta \rightarrow 0$.

The free parameters in Eq.~\eqref{sol4}, $\alpha, \beta$ and $w_0$, are obtained by fitting Eq.~\eqref{sol4} to time series describing the annual average GDP per capita of three different countries, each series scaled by its initial value.
The fitting procedure minimizes the sum of squares $E=\sum_i \log^2(w_i/d_i)$, where $i$ runs over the available data points.
\begin{table}
   \begin{center}
     \begin{tabular}{ |c|c|c|c|r|} 
    	\hline \hline
    	Country	& $w_0$   & $\alpha$ (yrs$^{-1}$)	& $\beta$ (yrs$^{-1}$) 	& $t^*$ (AD)	\\ \hline
    	UK		& 0.9670  & 	0.0014 		& 	2.807 \; 10$^{-11}$ & 2026		\\ \hline
    	France	& 0.7878     & 	 0.0017		& 	3.793 \; 10$^{-11}$  		& 2045		\\ \hline
    	Sweden	& 0.7718       &   0.0018  		&        10$^{-7}$         & 2023		\\
    	\hline \hline  
    	\end{tabular}
      \end{center}
    \caption{Columns 2-4 are the parameter values of Eq.~\eqref{sol4} obtained by minimizing
    the norm of the logarithmic difference between data and theoretical prediction. 
    Column 5 lists the dates at which the model wealth diverges. }
 \label{param}
\end{table}

\begin{figure}
 \vspace{-3cm}
\begin{tabular}{c}
 \vspace{-4cm}
      \includegraphics[width=0.9\linewidth]{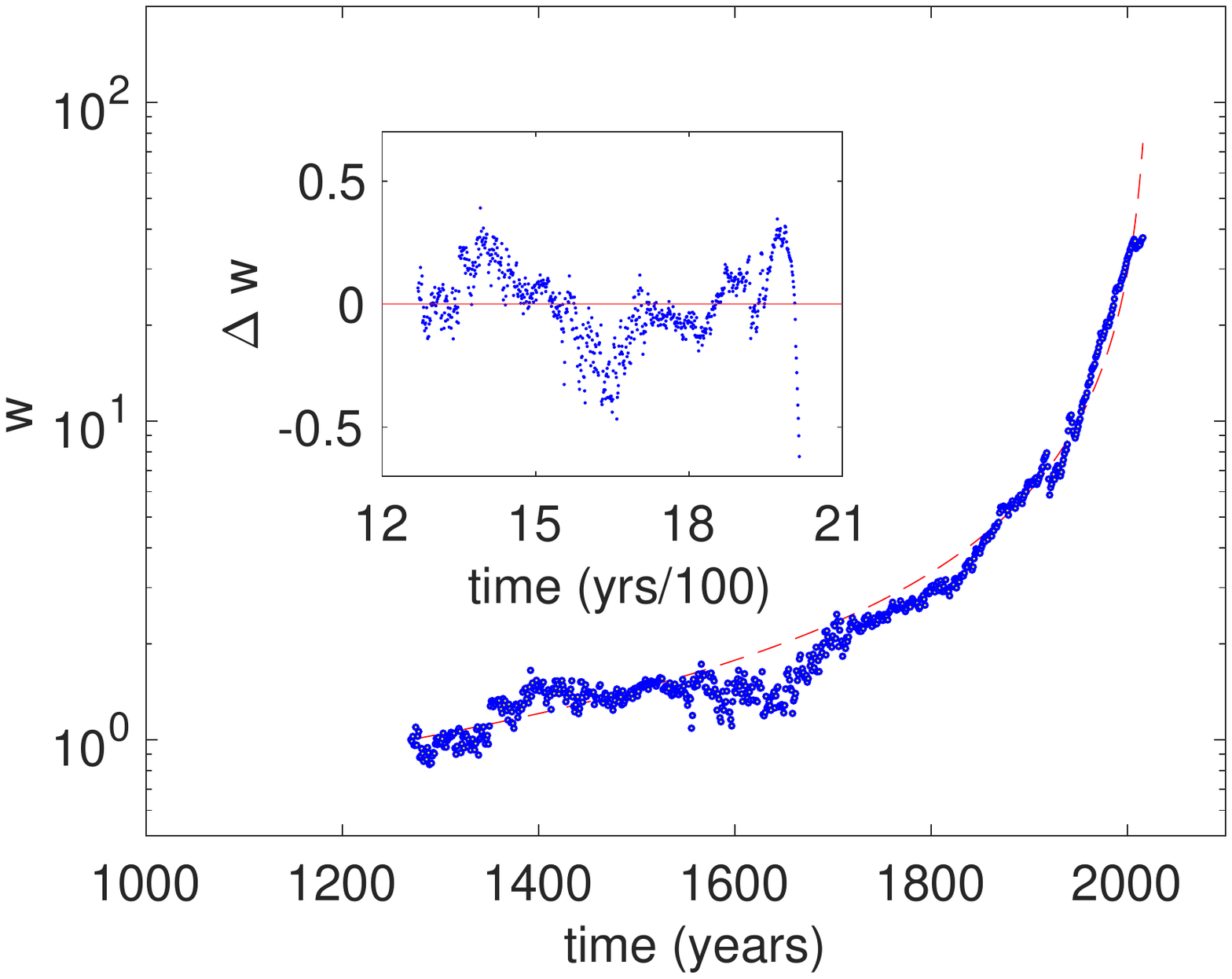} \\ 
       \vspace{-4cm}
      \includegraphics[width=0.9\linewidth]{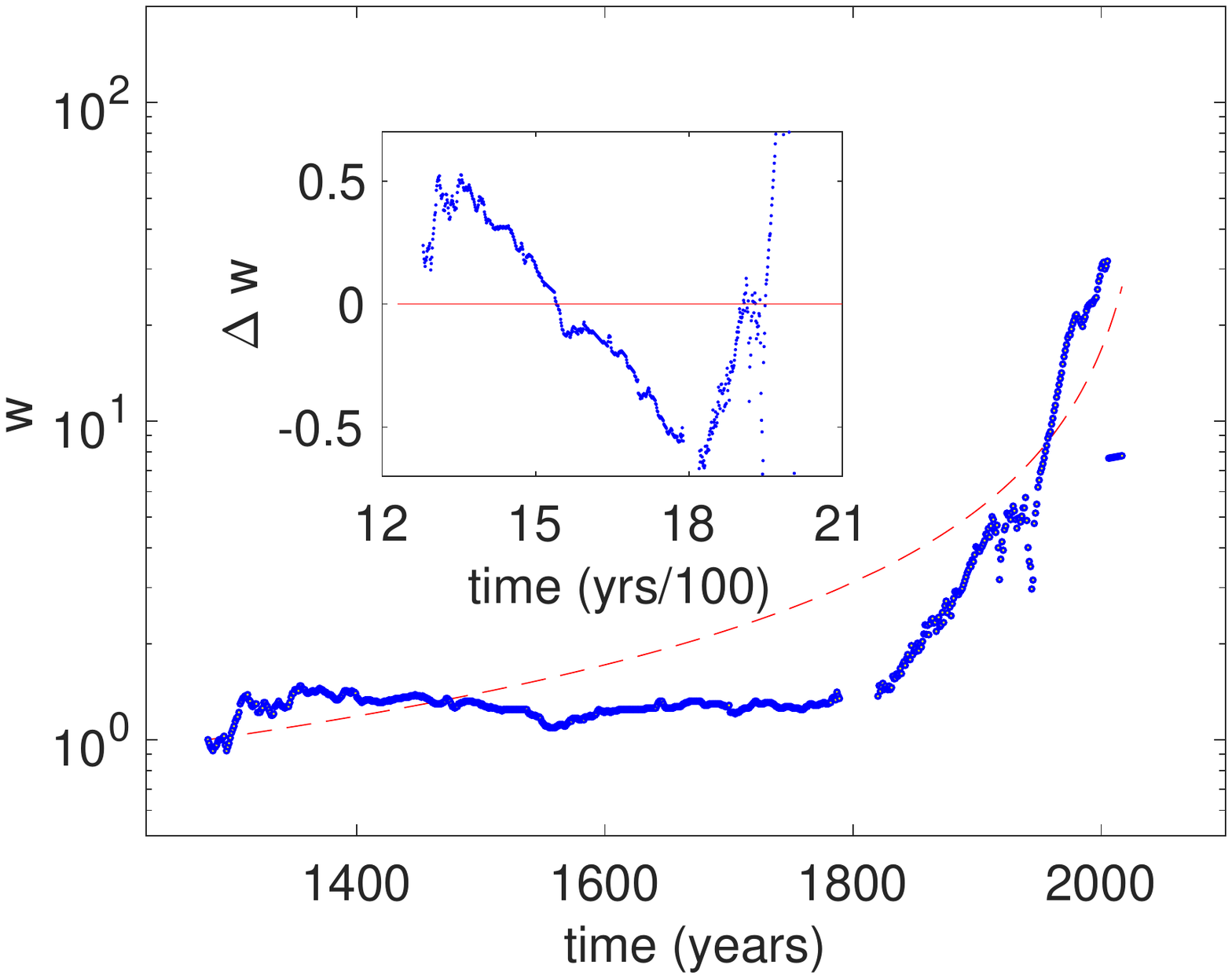} \\
       \vspace{-4cm}
      \includegraphics[width=0.9\linewidth]{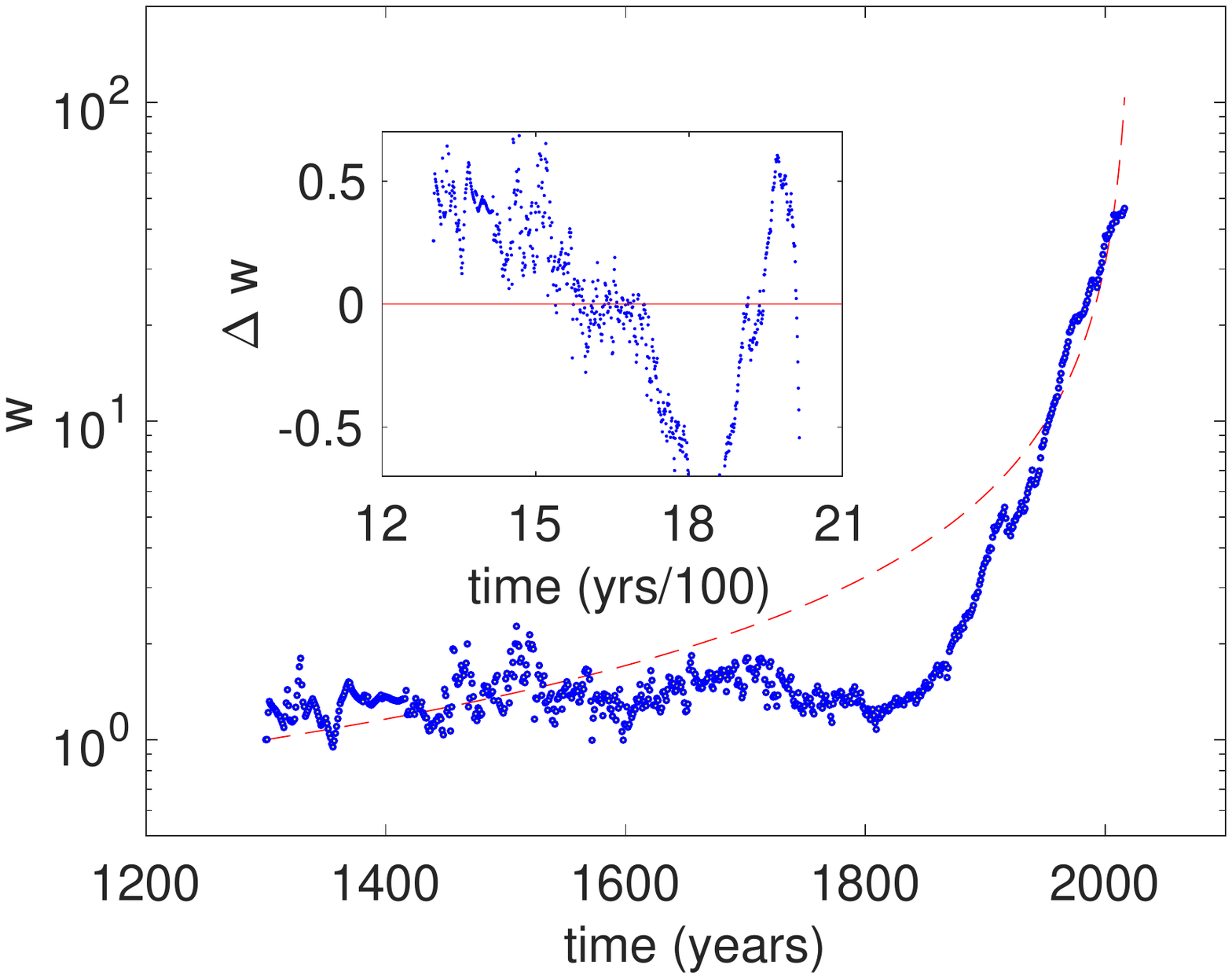}
	\end{tabular}
 \vspace{2cm}
     \caption{From top to bottom,   data and fitted trend are for the UK, France and Sweden, respectively.
     Time series of GDP per capita  are scaled by their initial value and plotted as blue dots on a logarithmic axis.
     A red hatched line depicts the fitted trend and the inserts show the de-trended data $\Delta w$ divided by
      100 for typographical convenience.
     The data fit utilizes Eq.~\ref{sol4} and the fit parameters are given in Table~\eqref{param}. 
       }
\label{fig:wealth1}
 \end{figure}
 In Fig.~\ref{fig:wealth1},  both GDP data (blue dots) and our predicted trend (red hatched line) are displayed for the UK, France and Sweden, respectively. 
In each panel, the insert shows the de-trended GDP fluctuations.

 Imagine now a scientist of yore who tried to fit the GDP data available at their  time using our present approach. 
 Would they   also predict a singularity occurring around   2025 AD?
 In other words, how far back in  time can the position of the singularity be predicted from  available data?
 
\begin{figure}
	\vspace{-2cm}
		 \includegraphics[width=0.9\linewidth]{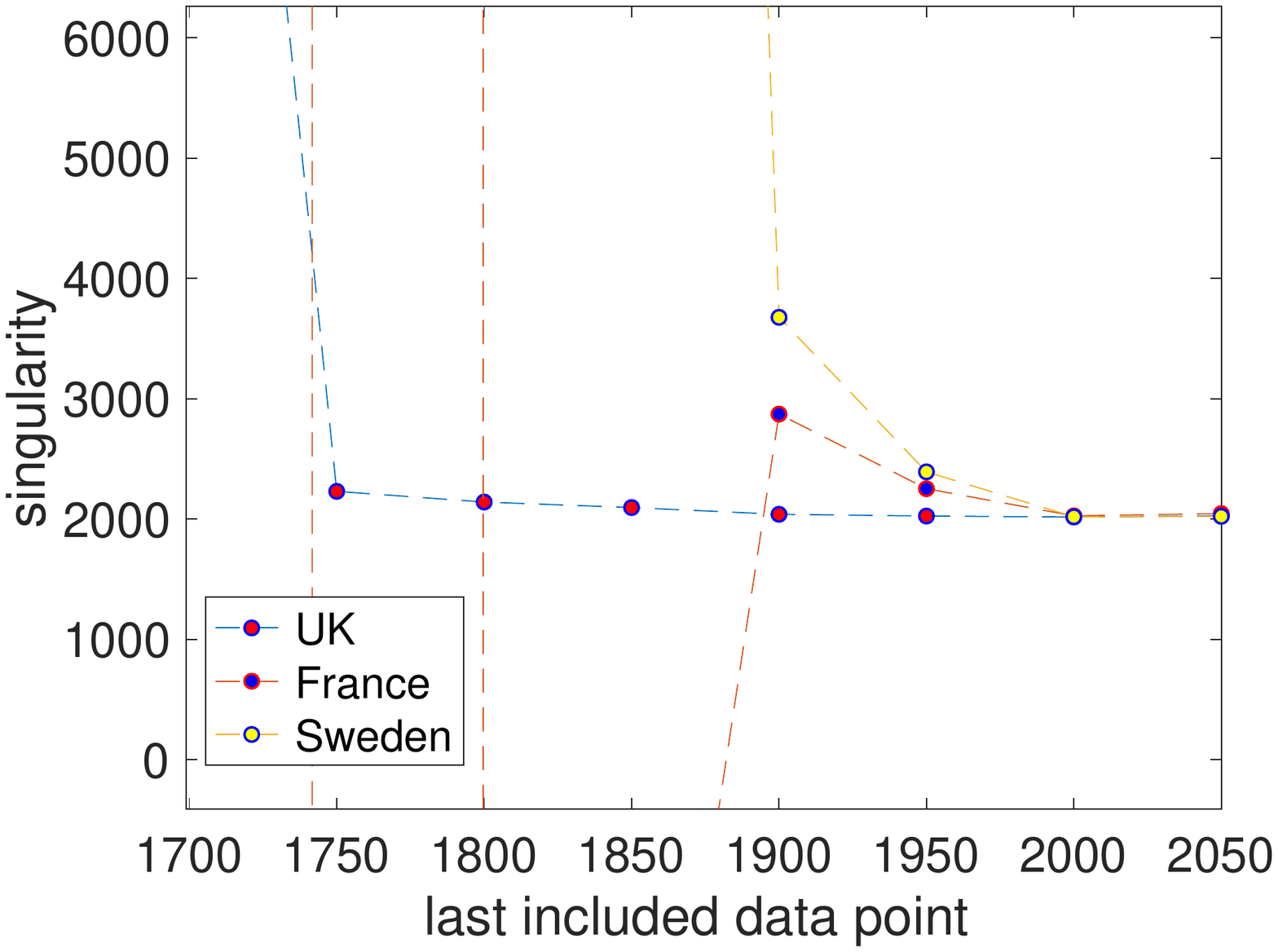} \\ 
		 \vspace{-2cm}
	\caption{For each country UK, France and Sweden, the date of the predicted singularity is plotted vs. the final
	year of a  curtailed GDP time series.
  }
\label{fig:sing_pred1}
\end{figure}

\begin{figure}
	\vspace{-3cm}
				 \includegraphics[width=0.9\linewidth]{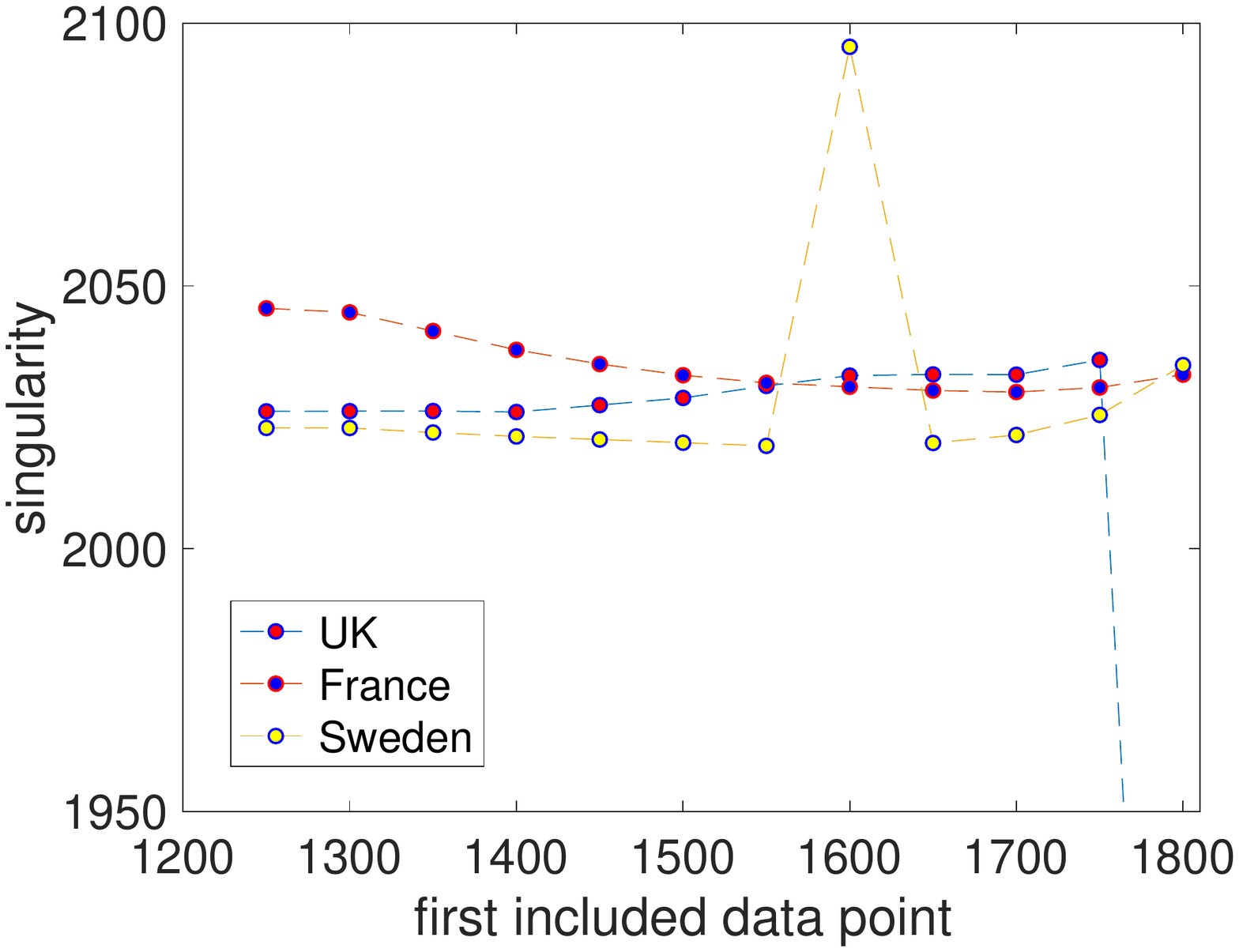} 
				  \vspace{-2.5cm}
	\caption{For each country UK, France and Sweden, the date of the predicted singularity is plotted vs.  the
	first year of a curtailed GDP time series. }
\label{fig:sing_pred2}
\end{figure}

 To investigate the issue, the later part of the full data set is curtailed, producing data streams ending at years 1700, 1750, 1800 AD, etc, and up to present days.
 Using the same algorithm on each curtailed data set  leads to the result shown in  Fig.~\ref{fig:sing_pred1}. 
 The UK data available in  1750 AD already yield a reasonable prediction, while 
 the French and Swedish data need to include times from
  from  1950 AD onwards. 
Conversely, to investigate the robustness of the results when only recent data are used,
   the initial parts of the data series can  be curtailed. The results are shown  in  Fig.~\ref{fig:sing_pred2}.
 Except for a single Swedish data point, the position of the singularity only shows minor dependence on the initial time of the series and
 for all three countries, series limited to the last four centuries already give consistent results.

\begin{figure}
 \vspace{-3cm}
\begin{tabular}{c}
 \vspace{-4cm}
      \includegraphics[width=0.9\linewidth]{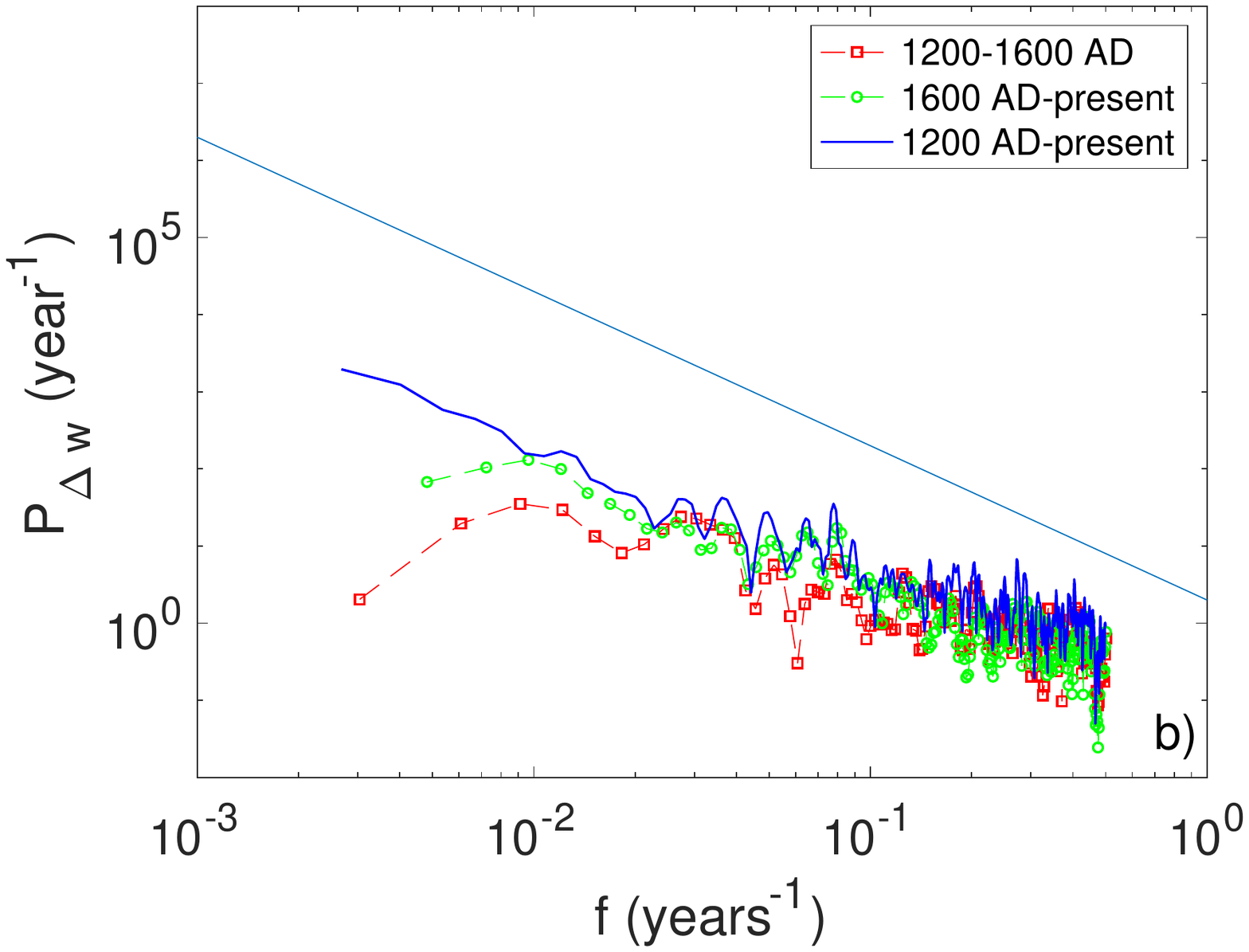}\\
     \vspace{-4cm}
      \includegraphics[width=0.9\linewidth]{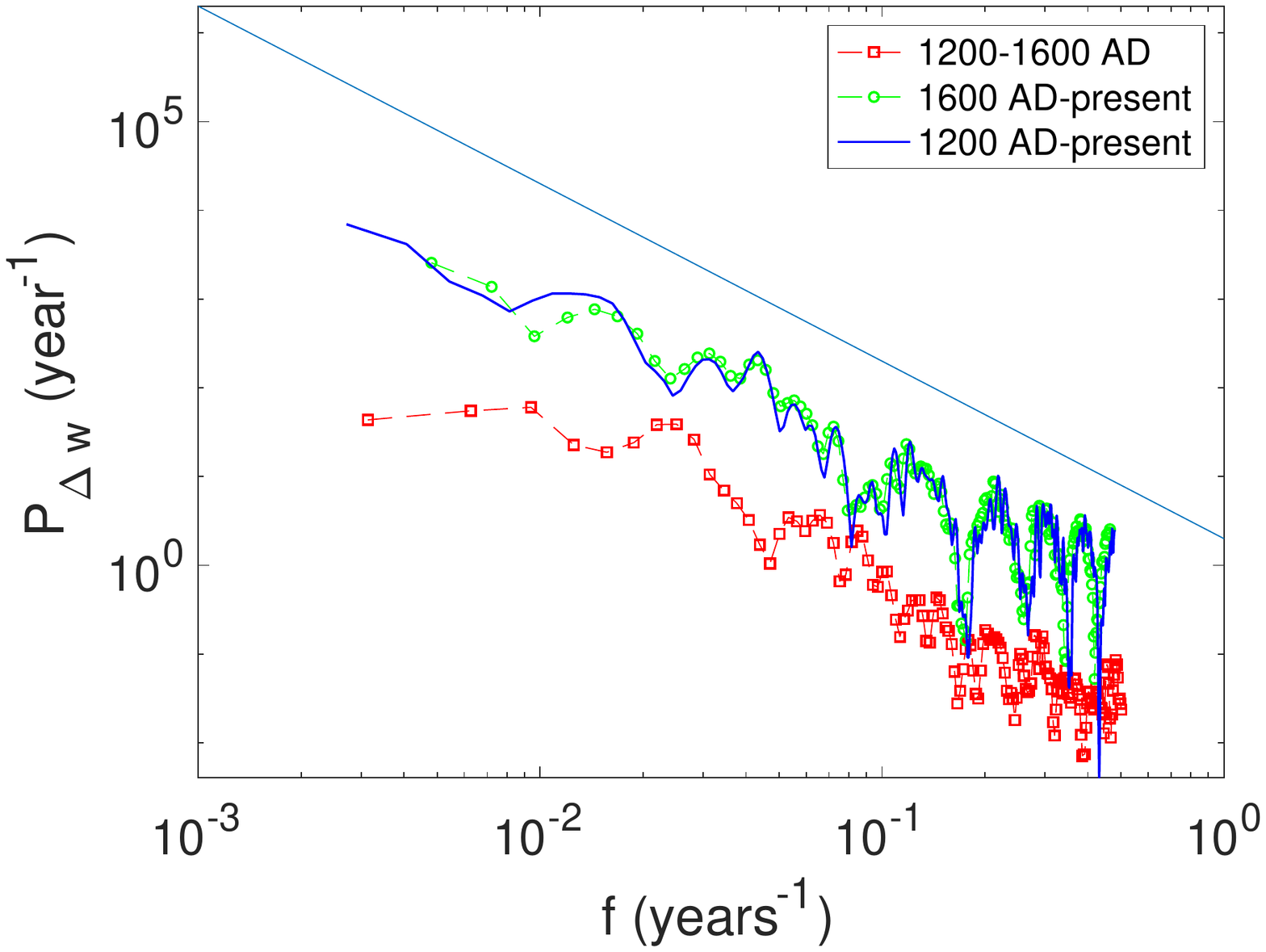}\\
     	\vspace{-4cm}
 \includegraphics[width=0.9\linewidth]{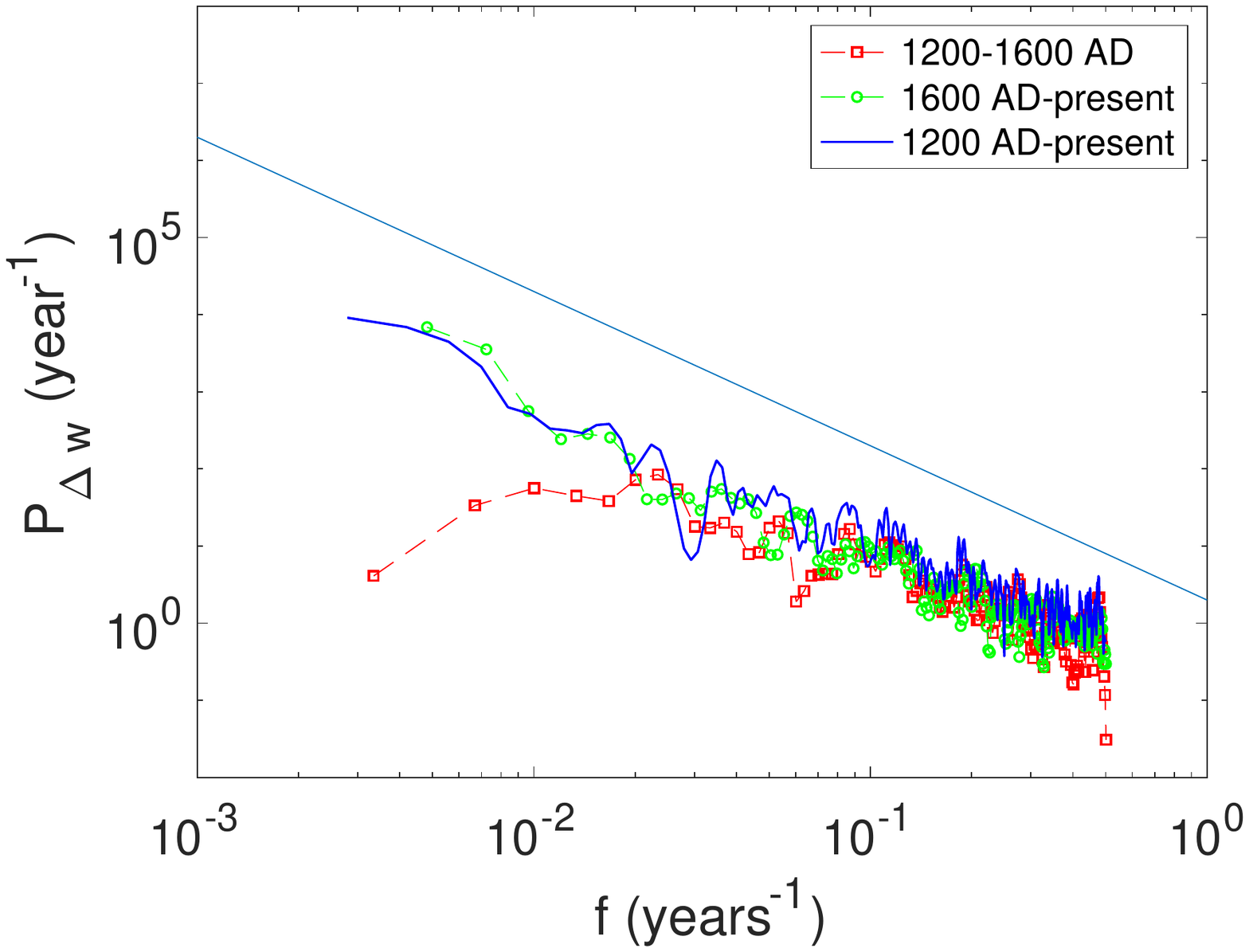}
	\end{tabular}
	 \vspace{1cm}
     \caption{From top to bottom, data  are for the UK, France and Sweden, respectively.
        For each country, three power spectra of the de-trended data $\Delta w$ are plotted vs. cyclic frequency. 
    As explained in the  legend, one set pertains to  the  full time series and the others to time series
    curtailed at the beginning and at the end of the period.
    The straight line is a guide to the eye, representing a power law with exponent $-2$.
     Curtailing has a minor effect on the position of the peaks and on the level of the
    $f^{-2}$ data trend. 
     }
\label{fig:wealth2}
 \end{figure}

Power spectra are usually extracted from stationary time series and 
 plotted as functions of the angular frequency $\omega$. In our case, the 
 stationarity of the  de-trended fluctuations
 can only be  approximate,  since the predicted trend diverges  and
 overshoots the data close to our present time. 
 Second,  the cyclic frequency
$f=\omega/2\pi$, i.e. inverse time, is used
 as independent variable for easier  identification  of the oscillation periods associated with the 
peaks seen in  Fig.~\ref{fig:wealth2}, where  power spectra are plotted for each country, from top to bottom
UK, France and Sweden.  In order to highlight historical  continuity,  three power spectra are extracted for each country
 from data pertaining to different  periods.
 
 Monotonically decreasing spectral intensities and uncertain small frequency behavior are generally  expected for 
economic time series~\cite{Granger66,Medel14}. In our case the same $ f^{-2}$  power law signature,
given by the full line, is shared by all spectra, irrespective of their epoch and country of origin. This power law signature was also  noted in shorter Danish GDP series~\cite{RasmussenMosekildeHolst1989}.

\begin{figure}
     \vspace{-3cm}
         	 \includegraphics[width=0.9\linewidth]{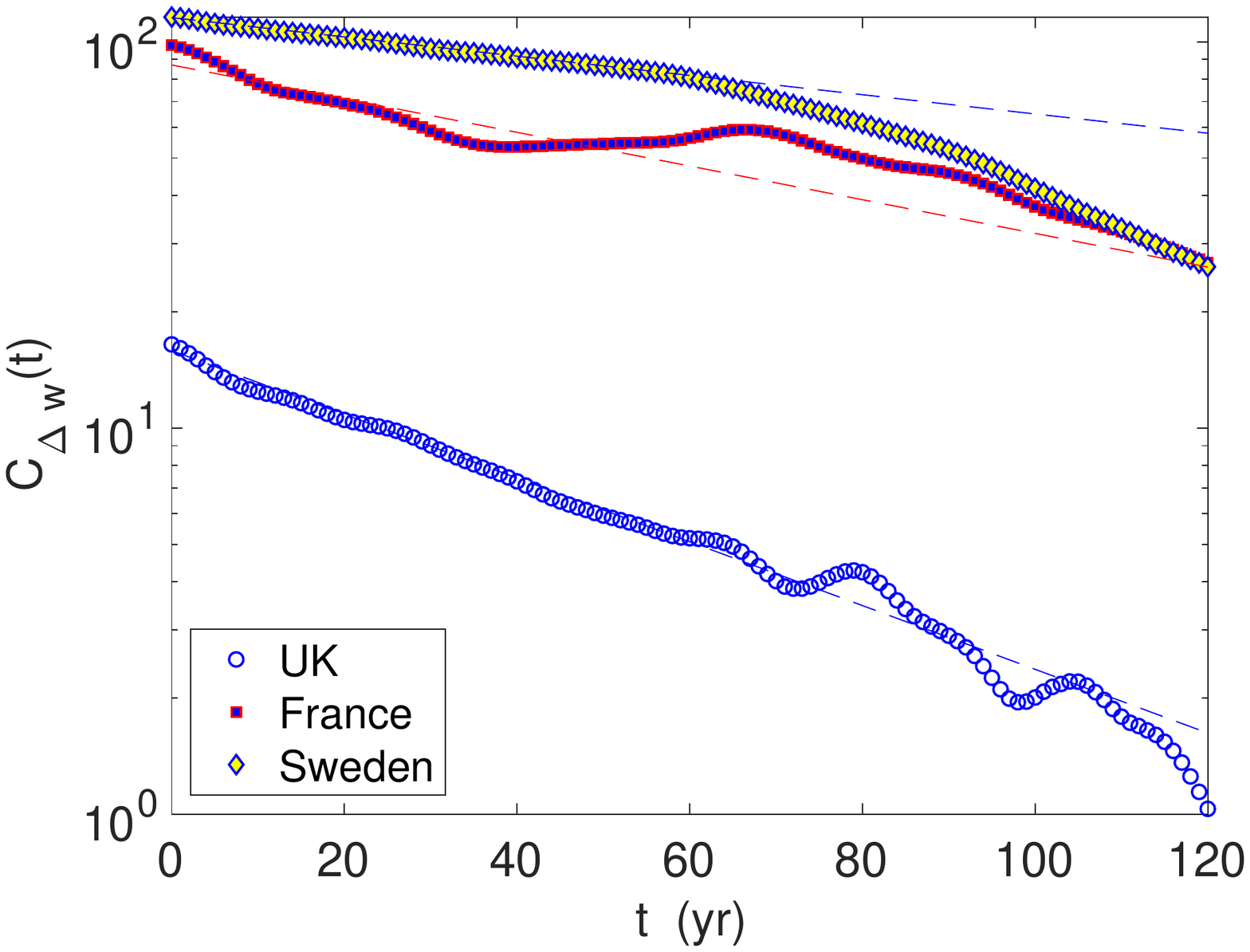}\\
     	     \vspace{-3cm}
     \caption{
     The autocorrelation function of the de-trended 
     data for the full series is plotted on a logarithmic 
     y axis. Initial exponential decays, marked with  hatched lines, are clearly  discernible.
     Their  decay times are estimated to be $T=53,\;100$ and  $174$ years for UK, French and Swedish data, respectively.
     }
\label{fig:3corrs}
\end{figure}

 To ensure  integrability, the background  contribution must level off, or possibly decay,  for $f \rightarrow 0$.
 If it levels off, it  is associated~\cite{Sibani13} with  an 
exponentially decaying  autocorrelation function of the time series of data.
Our time series are unfortunately too short to  accurately ascertain small $f$  behavior.
The presence of a  plateau for small $f$ is not clear, and the veering off of the signal can
also be produced by  a `long wave' contribution.
 As shown in  Fig.~\ref{fig:3corrs}, a straightforward inverse Fourier transform nevertheless yields an autocorrelation function with
a clear exponential decay at short lag times, with a decay time $T$ in the range $50-70$ years.

Oscillatory behavior is widely discussed in analyses of modern time series, see e.g.~\cite{Yegorov11,Korotayev10}. 
Cycles with a period of 7 - 11 years are usually referred to as business cycles, 
those with periods of 15 - 25 years are associated with  Simon Kuznets~\cite{Kuznets1930}, and
long cycles of 50 years or more with Nikolai Kondratiev~\cite{Kondratiev1925}. 
All these oscillations seem to be present in our data. 

\begin{figure}
     \vspace{-3cm}
         	 \includegraphics[width=0.9\linewidth]{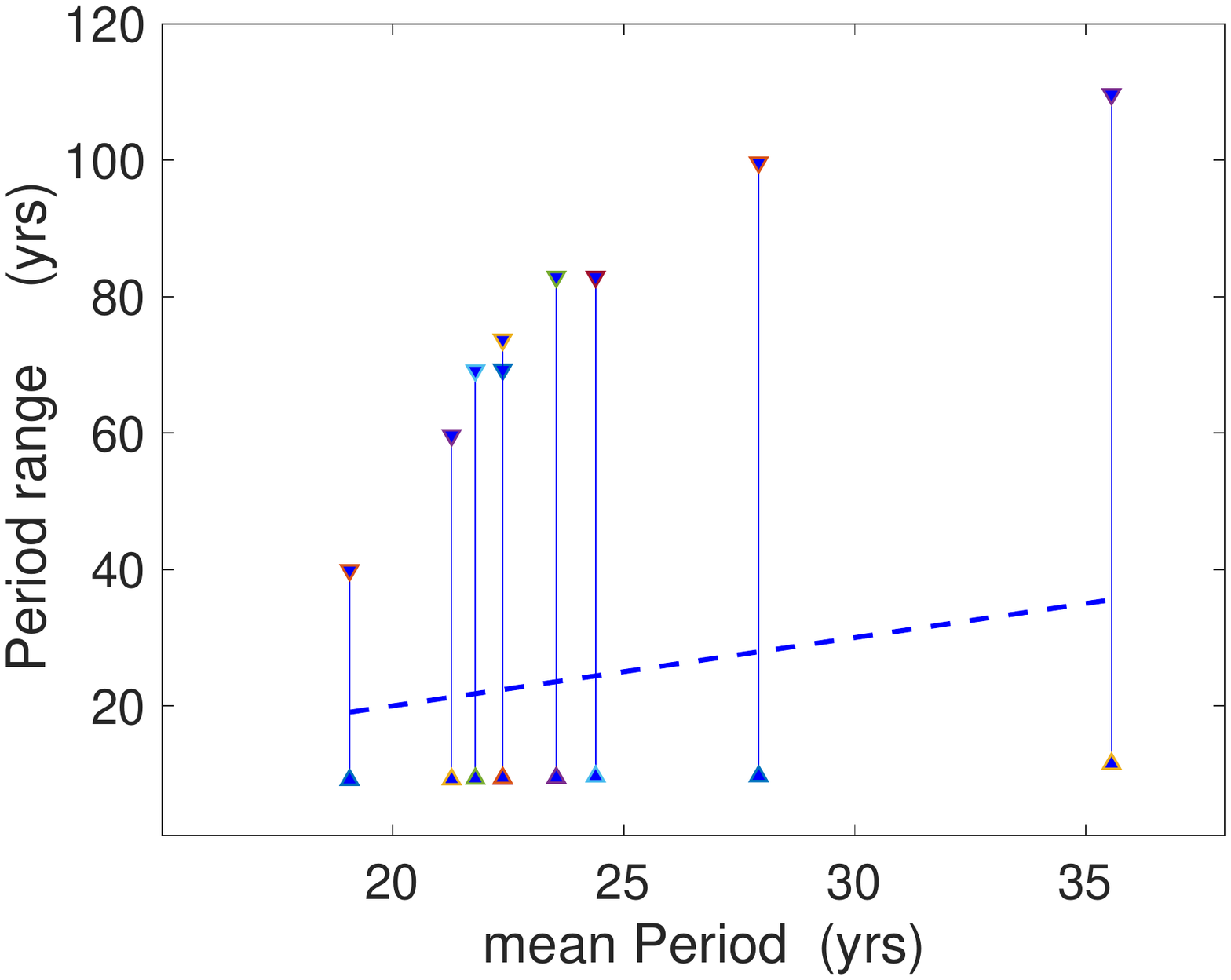} 
     	     \vspace{-3cm}
     \caption{
      For  each power spectrum, the period (inverse frequency) of the  ten  largest  peaks
     is  calculated, yielding a group of 
     nine different values for each peak index. 
     The range   of the values found in each  group is drawn  
     as  a vertical bar, and plotted vs the corresponding  mean.
     The hatched  line is a guide to the eye connecting  all  mean period values..  
     }
\label{fig:all_periods}
\end{figure}

The periods corresponding to the ten peaks of  lowest frequency are determined for all  spectra,
producing nine values  for each  peak index. These periods characterize oscillations
in  different countries and/or epochs. To investigate the variation across countries and epochs,
 the range of  values encountered  in  each group is  plotted  in 
Fig.~\ref{fig:all_periods}
 vs. the
corresponding mean period. Note that nearby groups have overlapping ranges, confirming  that the same or nearly the
same periods are found in data from different areas or epochs.
The   longest cycles feature   a stronger variation, which suggests caution in their interpretation.
  \vspace{-0cm}
  
 { 
 Tentatively, we associate  the form of the  power spectra 
  to  uncorrelated random events impinging on society, for short   `historic noise'.
  Historic noise both  triggers the   oscillatory behavior of   a number of coupled stakeholders and
  produces the background $f^{-2}$ signature. The latter
  describes  a lingering response
  to  a flow of  unpredictable events and the time 
      scale $T$ is what the  system typically  needs  to structurally re-adjust to perturbations,
  e.g. through  a slow change of  societal focus and/or structure.
   Such  hypothetical  scenario  generalizes
   descriptions only  including  cycles generated by  
 economic agents interacting in a market.
 That the  values of overall decay time 
are  close  to  the length of Kondratiev `long wave' cycles~\cite{Kondratiev1925} might  be coincidental, 
but a clarification certainly deserves  further analysis.
  }

\section{Discussion}
Our simplistic `single cause' model features a strong positive feed back loop: wealth growth is fundamentally generated by human interactions and part of the wealth produced goes into communication technologies that increase the amount of interactions per unit of time and thus human cooperation.  
This feedback loop is the model's basic assumption and produces its finite time singularity.

A recent study in cell phone use as a function of community size provides some anecdotal support for our key assumption. 
Using cell phone calls as indicators, increasingly frequent human interactions were recently linked to urbanization~\cite{Schlapfer14}.
The main finding is that the total number of contacts and the total communication activity grow super-linearly with city population size, implying that the larger the community is, the larger is the number interactions per capita. 
While not directly applicable to historical data, this result suggests that communication intensity and urban size were 
similarly related in the past.
Since human cultural development and urbanization proceeded in parallel as city sizes grew over time~\cite{Ritchie19},  communication intensity must  also have grown over time, which buttresses the assumption behind our model's finite time singularity. 

That wealth generation is assumed to be driven by the same mechanism over eight hundred years begs the question of which human circumstances have staid put through the dramatic changes which happened  in the same period, e.g. turning from  agricultural to industrial and to service based production.
White~\cite{White1967} and others~\cite{Ness1989} highlighted the strong emphasis that Western culture puts on the value of individuals together with their special relation to God and the rest of the world.
In the Jewish and Christian ethos these elements encourage individuals to pursue self-fulfilment and the achievement of the common good through a cooperative effort that involves communication among peers and the exploitation of available natural resources. 
The same principles permeate our moral codexes and generate deep-lying human interaction patterns, in particular the fact that wealth in part is used to increase human communication intensity.

Our model implicitly assumes the existence of a structured population with shared cultural values able to endogenously coordinate its resources for the achievement of common goals. 
The medieval church comes to mind as a relevant example from the past, and national states can partly fill this role for the present. 
In contrast, many geographical areas corresponding to modern national states do not fit well, having undergone considerable political re-structuring in the course of the last eight centuries.

Consider e.g. Italian GDP data, initially high compared to the rest of Europe, but remaining almost constant until the middle of the nineteenth century, when a phase of rapid economic development began.
Italian GDP  was initially sustained by financial flows linked to the commercial activities of e.g. the maritime republics, of other city states and of the Holy See.
These were centers of excellence and, in modern terminology, would count as international powers. 
The formation of a national Italian state is a more recent  development, roughly coinciding with the country's beginning industrialization.
These factors we believe explain the relatively bad fits obtained when applying the model  to Italy and Turkey, which seem to have abruptly imported an industrial society model during the 19th century. 
In contrast, the three countries chosen for our analysis maintained reasonably stable national and sociopolitical structures over the period considered, although Sweden's adoption of an industrial economy  also happened rather abruptly in the mid 19th century.
  
A different and well-established approach to explain historical GDP data posits that significant new human technology complexes, both physical -  e.g. the steam engine, the Internet, biotechnology -  and social - e.g. our laws, institutions, morals -  co-evolve and produce new exponential wealth-growth phases each with a characteristic exponent, and thus generate a new period of socio-technical development. 
Successive overlays of new significant technologies, each growing exponentially, do generate a super-exponential wealth-growth over an extended period ~\cite{MosekildeRasmussen1986}\cite{RasmussenMosekildeHolst1989}\cite{RosenlystSiboniRasmussen2018}, without reaching a finite time singularity.
In this framework, the abrupt shift in wealth development at the onset of the Industrial Age reflects significant societal and technological changes that warrant a different growth mechanism and thus a change of causal description. 
 {
Arguably,  the statistical properties of de-trended data can only retain a weak dependence on how the trend is extracted.
The oscillations emerging from our analysis have similar periods across countries and historical ages, and are all embedded in a stochastic process producing  the $f^{-2}$ decay of all power spectra.
This component does not seem to be a recognized feature of long GDP time series. 
While further work will be needed to ascertain its precise nature, a $f^{-2}$ decay in wealth fluctuation power spectra can hardly be ignored. 
}

 {
Finally, any type of time series analysis of partly constructed historical data as these should be cautioned with the methods by which the data are constructed. 
In our analysis we have assumed that neither a hyperbolic growth trend, a $f^{-2}$ decay in wealth development fluctuations, nor cyclic wealth components are used in the data construction. 
}

\section{Conclusion}
Time series covering the average GDP per capita through eight centuries for three European countries, the UK, France and Sweden,
 are used as proxies for cultural evolution, a term which includes both social and technological developments.
  The data are analyzed as superpositions of a historical trend and short term fluctuations. 
The trend is modelled using a simple ordinary differential equation (ODE), for `wealth', a dimensionless quantity
 which represents all societal features linked to economic activity. 
The  ODE is constructed in two steps.
Assuming that once generated wealth is never destroyed and using a 'natural' time variable $\tau$, produces a
 very simple, asymptotically exponential dynamics, i.e. a fixed yearly percentual increase, see Eq.~\eqref{def0}.
The second step introduces how the wealth generated directly enhances the collaborative infrastructure expressed 
by Eqs.~\eqref{naturalT} and ~\eqref{dtau/dt} resulting in Eq. ~\eqref{fullEq}.
This second step accelerates development in terms of 'wall clock' time, and produces
 a trend which grows faster than exponentially and culminates in a finite time singularity, basically happening at our present time.

The de-trended data are empirically described by their power-spectrum, expressed for convenience as a function of cyclic frequency $f$.
 {
A $f^{-2}$ background decay in power is seen in all investigated spectra and many peaks are vi\-sible 
with periods ranging  from about ten years up to the lifespan of individuals.
Though reminiscent of economic cycles, their presence in time series predating industrialization seems to require a more general explanatory mechanism.
Further work is required to clarify the mechanism behind these power spectra. }

 {
We propose  that random events, e.g. inventions, epidemics and wars drive both the noisy component
and  the oscillations of the GDP around its trend,
with a de-correlation time of the combined process of the order of half a century. 
This time scale exceeds the duration of one of the longest and bloodiest conflicts in Europe, the Thirty Years' War, 1618-1648.
Its  length might be related to that  of humans' active life, to the   memory span of individuals as well as to the life time of socio-technical
 structures. 
 }

\section{Outlook}
 Our  prediction that  GDP per capita will  soon  be infinite
is obviously unrealistic. Since the predicted
trend is nevertheless in reasonable agreement with data through many centuries,
structural changes must be under way. Many other observations, beside insights from historical data~\cite{Johansen01}, 
point to imminent and radical societal transformations 
 attributed e.g. 
to a growing mismatch between  new physical and social technologies~\cite{GrowingGap_2018}.
From our vantage point, this gap is the proximate cause of many current  disruptions,  e.g.  
human induced  climate change~\cite{ClimateChange} and the concurrent  ongoing sixth mass extinction~\cite{SixthMassExtinction}.
Their roots  date however much further back in time and stem from   established human interaction patterns. 

If deep-lying cultural themes condition  a development about to reach its limits, technological fixes alone 
will not solve the current predicament.  
Unfettered exploitation of our natural environment is a well entrenched  cultural trait,
already  framed as an expression of  Divine Will in the book of Genesis~\cite{Genesis}. 
This trait    has certainly fuelled  wealth
growth through many centuries, but its  limitations are increasingly  recognized in the
crowded world of today.
For example, Beinhocker and Hanauer~\cite{BeinhockerHanauer_2017} suggest
 to measure economic performance in terms of `its accumulated ability to solve human problems' 
 instead of in terms of `the market value of all produced products and services'.
This requires consensus on what  human problems are, but 
could give  increased political attention to
 climate change mitigation and to combating poverty and disease.
 
 A second historical trait is our unrelenting focus on
communication speed improvement. Together with the
ever increasing power of  information technologies, it
drives an evolutionary trend that upsets interaction patterns
established since time immemorial. 
Already a century ago, Pierre Teilhard de Chardin~\cite{Chardin59} 
imagined that a new evolutionary stage, called the `noosphere',
would inevitably emerge from the strong   interactions  of human minds.
  How to address a development in this general direction is  unclear, but with the obvious opportunities afforded by
interaction technologies, controlling its course  certainly demands major societal efforts.

 \subsection*{Acknowledgments}
 One of the authors, PS, would  like to thank Michael Christensen for his suggestion  to investigate  curtailed GDP time series.
 We are indebted to an anonymous referee for noticing an error in the submitted manuscript.
\bibliographystyle{apa-good}

\end{document}